\documentclass[aps,prl,twocolumn,noshowpacs,superscriptaddress]{revtex4}

\usepackage{CJK}
\usepackage{xspace}
\usepackage{multirow}
\usepackage{bm}
\usepackage{graphicx}
\usepackage{color}
\usepackage{amsmath,amssymb,amsfonts,amsthm}
\usepackage[colorlinks=true,linkcolor=blue,anchorcolor=red,citecolor=blue,urlcolor=blue]{hyperref}

\def \H {\mathcal{H}}

\begin{document}
\title{Index theorem on chiral Landau bands for topological fermions}

\author{Y. X. Zhao}
\email[]{zhaoyx@nju.edu.cn}
\affiliation{National Laboratory of Solid State Microstructures and Department of Physics, Nanjing University, Nanjing 210093, China}
\affiliation{Collaborative Innovation Center of Advanced Microstructures, Nanjing University, Nanjing 210093, China}
\author{Shengyuan A. Yang}
\address{Research Laboratory for Quantum Materials, Singapore University of Technology and Design, Singapore 487372, Singapore}

\begin{abstract}
Topological fermions as excitations from multi-degenerate Fermi points have been attracting increasing interests in condensed matter physics. They are characterized by topological charges, and magnetic fields are usually applied in experiments for their detection. Here we present an index theorem that reveals the intrinsic connection between the topological charge of a Fermi point and the in-gap modes in the Landau band structure. The proof is based on mapping fermions under magnetic fields to a topological insulator whose topological number is exactly the topological charge of the Fermi point. Our work lays a solid foundation for the study of intriguing magneto-response effects of topological fermions.
\end{abstract}

\maketitle

{\color{blue}\textit{Introduction}.}
Topological semimetals have been attracting tremendous research interest in recent years~\cite{Classification-RMP,Vishwanath-RMP}. In these states, novel topological fermions, like Weyl or Dirac fermions, emerge as excitations from multi-fold degenerate Fermi points.
A central concept here is that such Fermi points are characterized by topological charges in  momentum space, which play a key role in almost all interesting properties of these materials~\cite{Volovik:book,Vishwanath-RMP}.
For instance, the surface Fermi arcs of Weyl semimetals originate from the topological charges of the Weyl points (defined by Chern numbers)~\cite{WanXG2011prb}. Via identifying the topological charge, the classifications of topological gapless states have been made in the framework of $K$-theory for various symmetries~\cite{HoravaPRL05,ZhaoYXWang13prl,ZhaoWang16Aprprl,Classification-RMP}.

Investigating the response of a system under an applied magnetic field has been one of the most standard and powerful approaches in condensed matter research to extract system properties. For a three-dimensional (3D) system, the applied $B$ field quantizes the system's spectrum into Landau band structures along the field direction. Remarkably, it was found that for a Weyl point, its Landau band structure features a single \emph{gapless} chiral Landau band~\cite{nielsen1983adler}. This peculiar chiral Landau band is of fundamental importance. For example, it is at the heart of the chiral anomaly effect~\cite{Adler-axial,bell1969pcac,peskin1995quantum}, and results in a negative longitudinal magneto-resistance~\cite{nielsen1983adler}, which has been extensively studied in experiment~\cite{B_evidence-1,B_evidence-2,B_evidence-3,B_evidence-4,B_evidence-5,B_evidence-6,B_evidence-7,B_evidence-8,B_evidence-9,B_evidence-10}.

\emph{Is there a connection between the chiral Landau bands and the topological charges?} It is widely believed so. For the simplest case of a Weyl point, this connection can be made via a brute-force solution of the Landau spectrum, as in the original work by Nielsen and Ninomiya~\cite{nielsen1981absence,nielsen1983adler}. However, unlike in relativistic physics, the emergent fermions in condensed matter are not required to respect the Lorentz symmetry, and therefore may take much richer forms. For example, crystalline symmetries may not only lead to higher-order dispersions around the point~\cite{SM-1,SM-2,volovikLectNotes13,SM-3,SM-4,zhu2018quadratic,yu2019quadratic,wu2019higher},but also protect degeneracies (such as 3-, 4-, 6-, and 8-fold) beyond the simple twofold case for Weyl points~\cite{Bradlyn2016}. Some previous works have studied the magnetic response of a few specific cases, such as the multi-Weyl points~\cite{PhysRevB.96.085201,Yee2020} and triple points~\cite{Lepori2018}.  But, for the general cases, to the best of our knowledge, a proof of this important connection is still absent.


In this Letter, we fill this gap by proving the following index theorem:
\begin{equation}\label{IT}
  \nu=\mathcal{N},
\end{equation}
where $\nu$ and $\mathcal{N}$ are the topological charge for the zero-modes of the Landau bands and the topological charge of the Fermi point, respectively. In the case that  $\mathcal{N}$ is defined by the Chern number, $\nu$ is just the number of the chiral Landau bands (with its sign indicating the chirality).
Our proof goes by establishing a mapping from the target 3D semimetal under $B$ field to a 2D field-free \emph{semi-infinite} lattice. We show that the \emph{bulk} of the 2D lattice is insulating and carries the same topological number $\mathcal{N}$. Thereby, the zero-modes of the Landau bands exactly correspond to the gapless edge bands for the 2D topological insulator. Thus,  their topological charge $\nu$ is equal to $\mathcal{N}$, according to the faithful bulk-boundary correspondence of topological insulators~\cite{ZhaoYXWang14Septprb,Prodan-TI,Classification-RMP}. For the ease of presentation, we focus in the following on systems with topological charges defined by Chern numbers. The extension to other symmetry-protected topological fermions is straightforward and will be discussed at the end together with nontrivial examples. In addition, we show that the approach also gives a simple explanation for the zero-energy Landau levels in 2D systems such as single- or multi-layer graphene.


{\color{blue}\textit{Mapping from harmonic oscillator to semi-infinite lattice}.} Magnetic field quantizes the quasi-particle motion in the plane perpendicular to the field. It is well known that these in-plane degrees of freedom possess the same algebraic structure as a harmonic oscillator. Hence, we first introduce a mapping from a simple ``harmonic oscillator" to a lattice model.

A harmonic oscillator is described by the ladder operators, $a^{\dagger}$ and $a$, satisfying the canonical commutation relation, $[a,a^\dagger]=1$.  They operate on the Hilbert space spanned by the orthonormal particle-number basis $|n\rangle=(a^\dagger)^n|0\rangle/\sqrt{n!}$ with $n=0,1,2,\cdots$.
Their actions on these bases are defined by
\begin{equation}
\begin{split}
& a|0\rangle=0, \quad a|n\rangle=\sqrt{n}|n-1\rangle,~n>0,\\
& a^\dagger|n\rangle=\sqrt{n+1}|n+1\rangle,~~n=0,1,2,\cdots.
\end{split}
\end{equation}
And $N=a^\dagger a$ is the number operator, satisfying $N|n\rangle= n|n\rangle$.

\begin{figure}
	\includegraphics[scale=1]{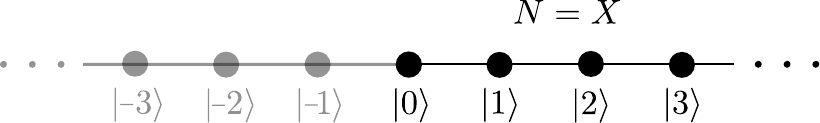}
	\caption{The Hilbert space of a semi-infinite lattice and its infinite extension. The Hilbert space of the semi-infinite $1$D lattice is equivalent to that of a harmonic oscillator, with the lattice-site position $X$ corresponding to the particle number $N$.\label{Lattice-extension}}
\end{figure}

A key observation is that the same algebra can be found for a lattice model defined on a 1D \emph{semi-infinite} lattice~\cite{Prodan-TI}. Here, we label the lattice sites by $n=0,1,2,\cdots$, and assume that each site has a single orbital basis denoted by $|n\rangle$ (see Fig.~\ref{Lattice-extension}). As a 1D lattice, we have the forward and backward lattice translation operators $T^\dagger$ and $T$, defined by
\begin{eqnarray}
&& T|0\rangle=0,\quad T|n\rangle=|n-1\rangle,~~n>0,\\
&& T^\dagger|n\rangle=|n+1\rangle ,~~n=0,1,2,\cdots.
\end{eqnarray}
We also have the position operator $X$, which reads off the the location of a basis:
\begin{equation}
X|n\rangle=n|n\rangle,~~n=0,1,2,\cdots.
\end{equation}
It is noteworthy that $T$ and $T^\dagger$ commute for every basis state except for the boundary $|0\rangle$, namely
\begin{equation}
[T,T^\dagger]=\delta_{0,X}.
\end{equation}

By comparing the above algebra, we identify an \emph{exact mapping} between the harmonic oscillator and the 1D semi-infinite lattice, with
\begin{equation}\label{mappings}
  a^\dagger=T^\dagger \sqrt{X+1},\quad a=\sqrt{X+1}\,T,\quad N=X.
\end{equation}
Inversely, we have
$
T^\dagger=a^\dagger \frac{1}{\sqrt{N+1}}$, $ T=\frac{1}{\sqrt{N+1}} a
$, and $X=N$. In the following discussion, the lattice and the spatial dimension $X$ emerging from our mapping will be termed as virtual lattice and virtual dimension, respectively.


Later in our proof, we shall utilize the bulk-boundary correspondence to argue the chiral edge bands from the topology of the \emph{bulk} virtual lattice. To this end, we will need to extend the semi-infinite lattice to be infinite, by adding bases $|-1\rangle,~|-2\rangle,~\cdots$, as illustrated in Fig.~\ref{Lattice-extension}.
On the infinite lattice, the two translation operators are denoted by ``adding a tilde'', and act as
\begin{equation}
\tilde{T}^\dagger|n\rangle=|n+1\rangle,\quad \tilde{T}|n\rangle=|n-1\rangle,
\end{equation}
with $n=\cdots,-2,-1,0,1,2,\cdots$. The extended operators are invertible and mutually independent, satisfying the relations
\begin{equation}
\tilde{T}^\dagger \tilde{T}=1,\quad [\tilde{T}^\dagger,\tilde{T}]=0.
\end{equation}

\begin{figure}
	\includegraphics[scale=0.25]{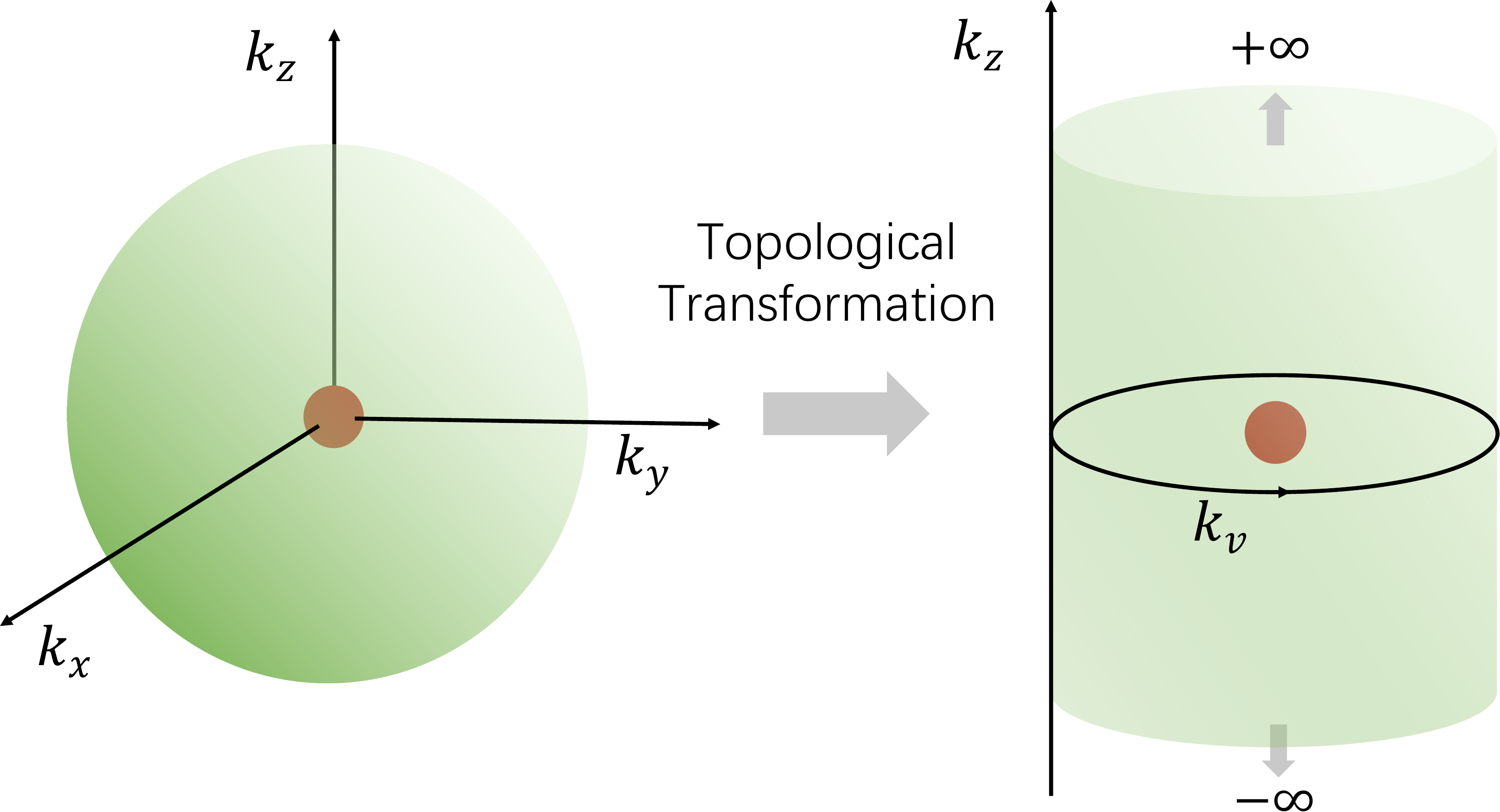}
	\caption{The transformation of the manifold that encloses a Fermi point. The topological charge of a Fermi point is defined on a $2$D sphere enclosing the Fermi point. After transformations, the resultant $2$D model with translational invariance has the momentum space being an infinite cylinder with the polar dimension parametrized by the virtual momentum $k_v$.\label{Base-manifold}}
\end{figure}

{\color{blue}\textit{A concrete example with chiral Dirac point}.}
Before presenting the general proof, it is helpful to illustrate our main ideas via a concrete model study.

Let us consider the following model for a four-fold degenerate chiral quadratic Dirac point,
\begin{equation}\label{Quadratic-Dirac}
\H=c_1(k_x^2-k_y^2)\Gamma_1+c_{2}2k_xk_y\Gamma_2+c_zk_z\mathrm{i}\Gamma_1\Gamma_2.
\end{equation}
Here the $c$'s are independent real model parameters, and $\Gamma_{\mu}$ with $\mu=1,2,\cdots,5$ are Dirac matrices~\cite{Dirac-Matrices}. This is a typical example of $k\cdot p$ effective model derived in solid state physics. And such Dirac point can be stabilized by crystalline symmetries, e.g., in space group No.~92~\cite{wu2019higher}.

Assuming the $c$'s are positive number, then the Fermi point described by Eq.~\eqref{Quadratic-Dirac} has the topological charge $\mathcal{N}=-4$. This topological charge can be evaluated by choosing a sphere $S^2$ to enclose the Fermi point in momentum space, on which the energy spectrum is gapped at zero energy. Thereby, $\mathcal{N}$ is given by the Chern number for valence bands restricted on $S^2$~\cite{Volovik:book,TKNN1982prl}:
\begin{equation}\label{Chern-charge}
\mathcal{N}=\frac{1}{2\pi}\int_{ S^2} d\theta d\phi~ \mathcal{F}_{\phi\theta}.
\end{equation}
Here, the Berry curvature $\mathcal{F}_{\phi\theta}=\partial_{\phi}\mathcal{A}_{\theta}-\partial_{\theta}\mathcal{A}_{\phi}$, with the Berry connection defined from the valence states $|\alpha,\bm{k}\rangle$ as  $\mathcal{A}_{\phi,\theta}=\sum_{\alpha}\langle \alpha,\bm{k}|i\partial_{\phi,\theta}| \alpha,\bm{k}\rangle$, $\theta$ and $\phi$ are the spherical angles for the momentum space.

Now, we exert a constant magnetic field along the $z$-direction. According to the minimal coupling principle, the field enters the model through the gauge covariant momentum operators
$\Pi_x=p_x-qA_x$ and $\Pi_y=p_y-qA_y$, with $B=\partial_x A_y-\partial_y A_x$. One can
linearly recombine the $\Pi$'s to form the bosonic ladder operators of a harmonic oscillator
\begin{equation}\label{particle-operators}
a^\dagger=\frac{\ell_B}{\sqrt{2}\hbar}(\Pi_x-i\Pi_y),\ \ \ a=\frac{\ell_B}{\sqrt{2}\hbar}(\Pi_x+i\Pi_y),
\end{equation}
where $\ell_B=\sqrt{\hbar/ qB}$ and $[a,a^\dagger]=1$. This is just the standard method to solve Hamiltonians with external fields, and the number operator $N=a^\dagger a$ gives the Landau index. For the effective model in Eq.~\eqref{Quadratic-Dirac}, the Hamiltonian for the Landau band structure is given by
\begin{equation}\label{TB-C4-B}
\H_B=\frac{1}{\ell_B^2}(\Gamma_- a^2+\mathrm{H.c.})+c_3 k_z \mathrm{i}\Gamma_1\Gamma_2,
\end{equation}
where $\Gamma_{-}=c_1\Gamma_1 -ic_{2}\Gamma_2$. Of course, for a given model, one can always solve the spectrum numerically and look for the chiral Landau bands. However, here, as a prelude to our general proof, we will demonstrate how to access the chiral Landau bands via the mapping we designed.


First, using the mappings in Eq.~\eqref{mappings}, we can transform Eq.~\eqref{TB-C4-B} to the following model on a semi-infinite virtual lattice,
\begin{equation}\label{TB-C4-SemiLattice}
{H}_L=\Big[\frac{1}{\ell_B^2}\Gamma_{-}(\sqrt{X+1}\ T)^2+\mathrm{H.c.}\Big]+c_3 k_z \mathrm{i}\Gamma_3\Gamma_4,
\end{equation}
which shares exactly the same spectrum (as the mapping is exact). Equation \eqref{TB-C4-SemiLattice} describes a $2$D system: the virtual dimension $X$ is represented in real space, while the physical $z$-dimension is in momentum space. One can see that in the bulk lattice, along the virtual dimension, the model has no lattice translational symmetry, since the hopping amplitude in Eq.~\eqref{TB-C4-SemiLattice} depends on the position of lattice sites. Another important observation is that, in the deep bulk for Eq.~\eqref{TB-C4-SemiLattice}, or equivalently, with sufficiently large Landau index $N(=X)=a^\dagger a$ for Eq.~\eqref{TB-C4-B}, the spectrum is gapped at zero energy. In other words, the 2D virtual lattice is insulating in the bulk. Thus, if the original Landau band structure possesses any gapless chiral bands, they must appear at the edge of our 2D semi-infinite lattice.

Our task is then to demonstrate that the bulk of our virtual lattice is actually a 2D topological insulator, such that gapless chiral edge bands must exist due to the bulk-boundary correspondence. For this purpose, we extend the semi-infinite lattice to be infinite for the virtual dimension, which gives
\begin{equation}\label{TB-C4-Lattice}
\tilde{H}_L=\Big[\frac{1}{\ell_B^2}\Gamma_{-}(\sqrt{|X|+1}\ \tilde{T})^2+\mathrm{H.c.}\Big]
+c_3 k_z \mathrm{i}\Gamma_1\Gamma_2.
\end{equation}
This is a $2$D insulator (as it should be), since $\sqrt{|X|+1}\tilde{T}$ is invertible and $\mathrm{i}\Gamma_1\Gamma_2$ anticommutes with $\Gamma_{1,2}$. We want to calculate the bulk invariant, i.e., the Chern number $\tilde{\mathcal{N}}$, of its valence states. Although the model has no translational symmetry along $X$, the bulk Chern number can still be evaluated via a real space method. Here, we proceed with an alternative way via adiabatic deformations. As we know, the Chern number is a topological invariant in the sense that it is constant under adiabatic deformations of the Hamiltonian, provided that the energy gap is not closed. Therefore, we can perform the smooth deformation $\sqrt{\lambda |X|+1}\tilde{T}$ for the hopping, by varying $\lambda\in[0,1]$, which preserves the invertibility of the operator and therefore the energy gap for the Hamiltonian \eqref{TB-C4-Lattice}. As $\lambda\rightarrow 0$, the resulting Hamiltonian is
\begin{equation}\label{TB-C4-Lattice-trans}
\tilde{H}_L=\Big(\frac{1}{\ell_B^2}\Gamma_{-}{\tilde{T}^2}+\mathrm{H.c.}\Big)+c_z k_z i\Gamma_1\Gamma_2,
\end{equation}
which acquires the lattice translational symmetry for the virtual dimension.
Performing the Fourier transform for the virtual dimension, we obtain the Hamiltonian in momentum space
\begin{equation}
\tilde{\H}_L=\Big(\frac{1}{\ell_B^2}\Gamma_{-}e^{\mathrm{i}2k_v}+\mathrm{H.c.}\Big)+c_z k_z i\Gamma_1\Gamma_2,
\end{equation}
where $k_v$ is the momentum for the virtual dimension.
Then, the Chern number $\tilde{\mathcal{N}}$ can be computed using the conventional formula, similar to Eq.~\eqref{Chern-charge}. But for the present case, the integration is over $k_v\in[-\pi,\pi)$ and $k_z\in(-\infty,\infty)$, namely, over the infinite cylinder $S^{1}\times \mathbb{R}$, as illustrated in Fig.~\ref{Base-manifold}.

For this particular model, straightforward calculations give $\tilde{\mathcal{N}}=-4$, which coincides with the topological charge $\mathcal{N}$ of the Fermi point. Consequently, there are four right-handed gapless chiral edge bands for the edge of the semi-infinite model, Eq.~\eqref{TB-C4-SemiLattice}. As we have argued, these gapless chiral edge bands just correspond to the chiral Landau bands in the Landau band structure of Eq.~\eqref{TB-C4-B}. Thus, the relation in (\ref{IT}) is confirmed.

Actually, as we shall see in the general proof below, the topological charge of the Fermi point is always equal to the Chern number of the resultant topological insulator, i.e., $\mathcal{N}=\tilde{\mathcal{N}}$ always holds. Moreover, it is worth noting that the index theorem actually ensures the topological robustness of the chiral Landau bands. Therefore, perturbations, for instance other terms in the $k\cdot p$ expansion, can be added to the Dirac point \eqref{Quadratic-Dirac}. As long as the Chern number is unchanged, the chiral Landau bands should stably persist.

{\color{blue}\textit{The index theorem}.} Now, we present a general proof of the proposed index theorem.
Consider the general form of a $k\cdot p$ model or a coarse-grained Hamiltonian for low-energy modes around a single Fermi point in a 3D system, (if the system has multiple Fermi points, we have a separate model for each point)
\begin{equation}\label{kp-model}
\mathcal{H}(\bm{k})=\sum_i g_i(\bm{k})M_i,
\end{equation}
where $M_i$ are a set of \emph{generic} constant Hermitian matrices with dimension corresponding to the degeneracy of the point (which can be arbitrary), accounting for the band degrees of freedom. As the model describes a single Fermi point, we assume that band crossings with zero
energy occur at and only at $\bm k=0$, where all $g_i$ are simultaneously equal to zero.
As illustrated in Fig.~\ref{Base-manifold}, we can always choose a sphere $S^2$ enclosing the Fermi point, on which the spectrum is gapped at the zero energy. The topological charge $\mathcal{N}$ of the Fermi point is given by the Chern number evaluated on this $S^2$, according to (\ref{Chern-charge}).

Without loss of generality, assume the applied magnetic field is along the $z$-direction. As aforementioned, by the minimal coupling principle, the Hamiltonian for the Landau band structure can be obtained by replacing $k_{x,y}$ with $\Pi_{x,y}$.
It is noted that $k_{x}$ and $k_y$ commute, whereas $[\Pi_{x},\Pi_{y}]=iB$. Hence, when implementing the replacements, one should be careful about the order of $k_x$ and $k_y$ in each polynomial $g_i(\bm{k})$, such that the Hermiticity of each term in Eq.~\eqref{kp-model} is preserved. By Eq.~\eqref{particle-operators}, $\Pi_{x,y}$ can be further expressed by the ladder operators, $a^\dagger$ and $a$. Thus, we arrive at the Hamiltonian,
\begin{equation}\label{Landau-Hamiltonian}
\mathcal{H}_B(a^\dagger,a,k_z)=\sum_i g_i(a^\dagger,a,k_z)M_i,
\end{equation}
where each $g_i(a^\dagger,a,k_z)$ is a Hermitian operator in terms of a polynomial of $a$ and $a^\dagger$, and parameterized by $k_z$.

Then, by our mapping in Eq.~\eqref{mappings}, $\mathcal{H}_B$ is equivalent to the Hamiltonian $H_L(T^\dagger,T, X,k_z)$ on the semi-infinite 2D virtual lattice, which can be further extended into $\tilde{H}_{L}(\tilde{T}^\dagger,\tilde{T}, |X|,k_z)$ defined on the infinite lattice. From the studied example, one observes that $\tilde{H}_{L}(\tilde{T}^\dagger,\tilde{T}, |X|,k_z)$ can be directly obtained from \eqref{Landau-Hamiltonian} by the transformations $a\mapsto \sqrt{|X|+1}\tilde{T}$ and its $\mathrm{H.c.} $, with $\tilde{T}$ and $\tilde{T}^\dagger$ the left and right unit translation operators for the virtual dimension. Recalling our assumption that all $g_i=0$ at $\bm k=0$, and otherwise \eqref{kp-model} has no zero-energy states, we see that $\tilde{H}_{L}$ has no zero-energy states and therefore is gapped, because $\sqrt{|X|+1}\tilde{T}$ is invertible. This means that $\tilde{H}_{L}$, the bulk of the semi-infinite lattice $H_L$, is an insulator. The possible chiral Landau bands in $\mathcal{H}_B$ must correspond to gapless chiral bands at the edge of $H_L$, which are in turn dictated by the bulk invariant $\tilde{\mathcal{N}}$ of the bulk model $\tilde{H}_{L}$ through the bulk-boundary correspondence:
\begin{equation}\label{bbc}
  \nu=\tilde{\mathcal{N}}.
\end{equation}

To evaluate $\tilde{\mathcal{N}}$, we perform the smooth deformation of $\sqrt{\lambda|X|+1} \tilde{T}$ with $\lambda\in [0,1]$, which preserves the invertibility. It follows that the Hamiltonian $\tilde{H}_{L}(\tilde{T}^\dagger,\tilde{T}, |X|,k_z)$ can be adiabatically deformed to $\tilde{H}_{L}(\tilde{T}^\dagger,\tilde{T},k_z)$, which eliminates the $|X|$ dependence and results in translational invariance for the virtual dimension.

Now, the Fourier transform, with $\tilde{T}\rightarrow e^{\mathrm{i}k_v}$, can be performed for the virtual dimension, which let us finally arrive at the Hamiltonian in momentum space $\tilde{\mathcal{H}}_L(k_v, k_z)$, with $k_{v}\in S^1$ parameterized as $[-\pi,\pi)$ and $k_z \in \mathbb{R}$.
By comparing the final result with the original model, one observes that the effect of the whole sequence of mappings and transformations can be
summarized by the transformation
\begin{equation}\label{Fundamental-substitution}
k_{\pm}\mapsto \frac{\sqrt{2}}{\ell_B}e^{\pm \mathrm{i}k_v},
\end{equation}
where $k_{\pm}=k_x\pm \mathrm{i}k_y$. Namely, $\tilde{\mathcal{H}}_L(k_v, k_z)$ can be directly obtained from the original model $\mathcal{H}$ in \eqref{kp-model} by the substitutions in (\ref{Fundamental-substitution}).

The correspondence in (\ref{Fundamental-substitution}) also offers us a shortcut to $\tilde{\mathcal{N}}$. As in Fig.~\ref{Base-manifold}, the topological charge of the Fermi point is evaluated by integrating the Berry curvature over the sphere $S^2$. In spherical coordinates, $k_{\pm}=R\sin\theta e^{\pm i\phi}$ and $k_z=R\cos\theta$. From (\ref{Fundamental-substitution}), one observes that $k_v$ just corresponds to the azimuthal angle $\phi$ in the original model. Thus, the Chern number $\tilde{\mathcal{N}}$ evaluated for the whole $k_v$-$k_z$ manifold for $\tilde{\mathcal{H}}_L(k_v, k_z)$ is the same as that evaluated on an infinite cylinder $S^1\times\mathbb{R}$ for the original Fermi point model $\mathcal{H}$. Then, one immediately realizes that \begin{equation}\label{NN}
  \tilde{\mathcal{N}}=\mathcal{N},
\end{equation}
because exactly the same topological charge is enclosed by both  $S^2$ and $S^1\times\mathbb{R}$. In other words, from the Fermi point to the resultant insulator, only the base manifold adopted to collect the total Berry flux emitted from the monopole is changed from $S^2$ to $S^1\times\mathbb{R}$. Therefore, the two Chern numbers must be the same.

Combining (\ref{bbc}) and (\ref{NN}), we close the proof and arrive at the index theorem in Eq.~(\ref{IT}).

{\color{blue}\textit{Discussion}.}
In the proof, we have focused on the topological charge defined by the Chern number, which is the most interesting and most frequently encountered case in condensed matter, as it does not rely on any symmetry. There also exist topological charges of Fermi points depending on certain symmetries.
Our index theorem applies to such symmetry-protected topological charges as well, as long as the magnetic field preserves the protecting symmetry. This can be readily understood by noting that if the symmetry is preserved, the topological charge must be maintained through the mapping (which only changes the shape of the enclosing 2D sub-manifold, as illustrated in Fig.~\ref{Base-manifold}).

For example, we know magnetic field does not break the sublattice (chiral) symmetry. Consider a 3D Dirac point, $\mathcal{H}=\bm k\cdot\bm\Gamma$, protected by the sublattice symmetry $\mathcal{S}=\Gamma_5$, and the mirror symmetry $M_z=\Gamma_3\Gamma_4 I_z$ with $I_z$ the inversion of the $z$ coordinate~\cite{Dirac-Matrices}.
These symmetries protect a $\mathbb{Z}$-valued topological charge~\cite{Chiral_mirror_charge}, and the Fermi point here has a unit charge. The magnetic field along $z$ preserves the mirror symmetry $M_z$ and the sublattice symmetry $\mathcal{S}$. The corresponding lattice model obtain by our transformation \eqref{Fundamental-substitution} is $\mathcal{H}_{L}=\frac{\sqrt{2}}{\ell_B}(e^{\mathrm{i}k_v}\Gamma_{-}+\mathrm{H.c.})+k_3\Gamma_3$, with $\Gamma_{\pm}=(\Gamma_{1}\pm \mathrm{i}\Gamma_2)/2$. The resultant 2D topological insulator carries the same symmetry-protected topological number, which dictates a pair of helical edge bands. According to our index theorem, this means there also exist helical bands in the Landau band structure, corresponding to the topological charge of the Fermi point.

Our analysis can also be adapted to 2D. For example, it is well known that the $2$D Dirac fermions described by
\begin{equation}
\mathcal{H}=v(k_-\sigma_++k_+\sigma_-),
\end{equation}
have the topological charge $\mathcal{N}=1$ protected by the chiral symmetry $\sigma_3$. Following the procedure of our general theory, the transformation \eqref{Fundamental-substitution} gives us the corresponding $1$D lattice model $\tilde{\mathcal{H}}_L=\frac{\sqrt{2}v}{\ell_B}e^{-ik_v}\sigma_++\mathrm{H.c.}$, which just corresponds to the famous SSH model~\cite{SSH-model}, and therefore has the bulk topological number $\tilde{\mathcal{N}}=1$. The zero-energy end mode of the SSH model is exactly the zero-energy Landau level in the Landau spectrum for the Dirac fermions. This argument directly generalizes to $\mathcal{H}_{n}=v(k_-^n\sigma_++k_+^n\sigma_-)$ with $n>1$, for which there must be $n$ zero-energy Landau levels. Interestingly, such $2$D Dirac point can be realized in chirally stacked multi-layer graphene~\cite{volovikLectNotes13,Multilayer-graphene}, for which such zero-energy Landau levels have been obtained via direct calculations. Our current analysis reveals their intrinsic connection to the topological charges.

Our theorem directly applies to real materials with topological nodal points, such as the TaAs-family Weyl semimetals~\cite{Weng2015,Huang2015}, HgCr$_2$Se$_4$~\cite{SM-1}, CoSi-family compounds~\cite{Tang2017}, and YRu$_4$B$_4$~\cite{wu2019higher}. The predicted chiral Landau bands can be imaged by using Landau level spectroscopy~\cite{Jeon2014} or magnetoinfrared spectroscopy~\cite{Chen2015e}.

\begin{acknowledgements}
{\color{blue}\textit{Acknowledgements.}} This work is supported by the Fundamental Research Funds for the Central Universities (Grant No. 14380119), National Natural Science Foundation of China (Grant No. 11874201), and the Singapore Ministry of Education AcRF Tier 2 (MOE2017-T2-2-108).
\end{acknowledgements}

\bibliographystyle{apsrev}
\bibliography{Ref_Landau_Index}

\begin{thebibliography}{46}
\bibitem[{Dir()}]{Dirac-Matrices}
\bibinfo{note}{The Dirac matrices $\Gamma_\mu$ with $\mu=1,2,\cdots,5$ are
  Hermitian and satisfy $\{\Gamma_\mu,\Gamma_\nu\}=2\delta_{\mu\nu}1_4$. All
  representations are equivalent up to unitary transformations. For instance,
  we can choose them as $\Gamma^\mu$ $\sigma_0\otimes\tau_1$,
  $\sigma_3\otimes\tau_2$, $\sigma_0\otimes\tau_3$, $\sigma_1\otimes\tau_2$,
  $\sigma_2\otimes\tau_2$.}
\bibitem[{Chi()}]{Chiral_mirror_charge}
\bibinfo{note}{Because of the mirror symmetry, the topological charge on $S^2$
  can be reduced to be on the reflection-invariant equator $S^1$. Then, the
  chiral symmetry leads to a $\mathbb{Z}$-valued topological charge.}
\end{thebibliography}

\end{document}